\documentclass{eas}

\usepackage{graphicx}

\begin{document}

\TitreGlobal{Mass Profiles and Shapes of Cosmological Structures}

\title{Gravitational Lensing \& Stellar Dynamics}

\author{Koopmans,~L.V.E.}\address{Kapteyn Astronomical Institute,
P.O.Box 800, 9700\,AV Groningen, The Netherlands}
\runningtitle{Gravitational Lensing \& Stellar Dynamics}


\begin{abstract}  
Strong gravitational lensing and stellar dynamics provide two
complementary and orthogonal constraints on the density profiles of
galaxies. Based on spherically symmetric, scale-free, mass models, it
is shown that the combination of both techniques is powerful in
breaking the mass-sheet and mass-anisotropy degeneracies. Second, 
observational results are presented from the Lenses Structure \& Dynamics
(LSD) Survey and the Sloan Lens ACS (SLACS) Survey collaborations to
illustrate this new methodology in constraining the dark and stellar
density profiles, and mass structure, of early-type galaxies to
redshifts of unity.
\end{abstract}

\maketitle

\section{Introduction}
Strong gravitational lensing has matured over the last decades in to a
powerful tool in astrophysical and cosmological studies. Whereas
research initially focused on cosmology, in particular on the
determination of cosmological parameters, and conducting large surveys
to find new strong lens systems, emphasis is now shifting toward the
use of strong lenses in the study of the mass distribution of galaxies
and their environments (see e.g.\ Kochanek et al.\ 2004). This became
feasible in the 1990s because of the advent of powerful new space and
ground-based telescopes for imaging and spectroscopic follow-up of
many newly discovered lens systems.

The combination of high-resolution multi-colour optical imaging with
spectroscopic redshifts, and sometimes with high-resolution
radio-imaging and more recently X-ray observations, of these lens
systems, is now providing a very detailed picture of the lens galaxies
and allows their mass, some-times mass profile, to be constrained.
However the mass-profile degeneracy introduces a limit to the accuracy
with which mass-profiles of strong-lens galaxies can be determined.

On way to break this degeneracy is to have {\sl independent}
information on the lens mass distribution, on a scale different
from the Einstein radius. One particularly powerful measure of galaxy
mass, or its density profiles, is obtained through stellar kinematics.
The latter has been particularly productive in the study of the mass
distribution in nearby galaxies. However, the high-quality data
required to break for example the mass-anisotropy degeneracy has severely
limited similar studies of distant galaxies, even with the availability of
8--10\,m class telescopes.

The combination of strong gravitational lensing and stellar-kinematic
constraints -- the focus of this review -- provides a way to break the
above-mentioned degeneracies in both methods {\sl and} overcome the
additional limit set by the high S/N-ratio required for the
stellar-kinematic data. In Section 2, using scale-free spherical
models, it is shown how the combination of lensing and dynamics can
break existing model degeneracies. In Section 3, the results from two
observational programs (LSD and SLACS) are highlighted regarding
constraints on the dark and stellar mass structure of early-type
galaxies to $z\approx 1$. In Section 4, a short summary and 
forward look is given.

\section{Spherical Scale-Free Lensing \& Stellar-Dynamical Models}

Despite the complexity of real galaxies, studying spherically
symmetric scale-free galaxies turns out to be enlightening in touching
upon the strengths and weaknesses of combining gravitational lensing and
stellar-kinematic constraints.

\subsection{Combining Lensing Mass and Stellar Velocity Dispersions}

Let us suppose the following spherical scale-free model for the lens 
galaxy:
\begin{equation}
  \left\{
  \begin{array}{lcl}
    \nu_l(r)    & = & \nu_{l,0} r^{-\delta} \medskip \\
    \nu_\rho(r) & = & \nu_{\rho,0} r^{-\gamma'}\medskip \\
    \beta(r)    & = & 1-{\langle\sigma^2_\theta\rangle}/
    {\langle \sigma^2_r} \rangle
  \end{array}
  \right.,
\end{equation}
where $\nu_l(r)$ is the luminosity density of stars -- a trace
component -- embedded in a total (i.e.\ luminous plus dark-matter)
mass distribution with a density $\nu_\rho(r)$. The anisotropy of the
stellar velocity ellipsoid is $\beta$, constant with radius. For a
lens galaxy with a projected mass $M_{\rm E}$ inside the Einstein
radius $R_{\rm E}$, the luminosity weighted average line-of-sight
velocity dispersion inside an aperture $R_{\rm A}$ is given, after
solving the spherical Jeans equations, by
\begin{equation}\label{eq:sigav}
  \langle \sigma_{||}^2\rangle (\le R_{\rm A}) = \frac{1}{\pi}
  \left[ \frac{G M_{\rm E}}{R_{\rm E}} \right] f(\gamma',
  \delta,\beta) \times \left(\frac{R_{\rm A}}{R_{\rm
  E}}\right)^{2-\gamma'},
\end{equation}
where
\begin{eqnarray}
  f(\gamma', \delta,\beta) = 2 \sqrt{\pi}\,\left(\frac{\delta -3}{(\xi
  - 3)(\xi - 2\beta)} \right)&\times&
  \left\{\frac{\Gamma[(\xi-1)/2]}{\Gamma[\xi/2]} - \beta
  \frac{\Gamma[(\xi+1)/2]}{\Gamma[(\xi+2)/2]} \right\} \times
  \nonumber\\ && \left\{\frac{\Gamma[\delta/2]\Gamma[\gamma'/2]}
  {\Gamma[(\delta-1)/2]\Gamma[(\gamma'-1)/2)]}\right\}
\end{eqnarray}
with $\xi = \gamma'+\delta -2$.  Similarly,
\begin{equation}\label{eq:sigr}
  \sigma_{||}^2(R) = \frac{1}{\pi} \left[ \frac{G M_{\rm
  E}}{R_{\rm E}} \right] \left(\frac{\xi - 3}{\delta -3}
  \right) f(\gamma', \delta,\beta) \times \left(\frac{R}{R_{\rm
  E}}\right)^{2-\gamma'}.
\end{equation}
In the simple case of a SIS with $\gamma'=\delta=\xi=2$ and
$\beta=0$, we recover the well-known result
\begin{equation}
  \sigma_{||}^2(R) = \frac{1}{\pi} \left[ \frac{G M_{\rm
  E}}{R_{\rm E}} \right]~~~({\rm SIS}),
\end{equation}
for the Singular Isothermal Sphere mass model. At high redshifts
(e.g.\ $z\sim 1$), one can often only hope to measure a
luminosity-weighted stellar velocity dispersion of a (lens) galaxy
within a give aperture, even when observed with 8-10\,m class
telescopes. In combination, with the mass measured inside the Einstein
radius of a lens galaxy, how does this provide us with information
about the galaxy's density profile? When studying Equation
\ref{eq:sigav}, we see two ways $\langle \sigma_{||}^2\rangle$ is
changed: through the exponent ($2-\gamma'$) and
through a change in the normalisation~$f(\cdots)$. 
\begin{itemize}

    \item If $R_{\rm A}$=$R_{\rm E}$, the normalisation only
      depends on~$\gamma'$. For a given $M_{\rm E}$, the slope
      $\gamma'$ then determines the gradient of the potential inside
      that radius. Steeper density profiles lead to larger potential
      gradients and, on-average, larger stellar velocities for a
      given luminosity profile ($\delta$) and given anisotropy
      $\beta$.

    \item If $R_{\rm A}$$\neq$$R_{\rm E}$, the stellar velocity
      dispersion inside an aperture also depends on $\gamma'$ through
      the factor $\left({R_{\rm A}}/{R_{\rm
      E}}\right)^{2-\gamma'}$. Hence, roughly speaking, gravitational
      lensing determines the mass inside $R_{\rm E}$ and stellar
      kinematics determining the mass inside $R_{\rm A}$. The average
      density profiles between $R_{\rm E}$ and $R_{\rm A}$ is then
      quantified, with the stellar velocity anisotropy often a
      secondary effect.

\end{itemize}

Both physical interpretations are at the basis of determining the average
density profile in the inner regions of gravitational-lens
galaxies. Real galaxies are obviously more complex, but this
toy-model illustrates the basic ideas of combining lensing and
dynamics and how it is used to measure the density profiles of (lens)
galaxies and break the mass-profile and mass-anisotropy degeneracies.

\subsection{How reliable can the density profile be measured?}

To answer this question, we take the derivative of
Equation~\ref{eq:sigav} with respect to $\sigma_{||}$ and $M_{\rm E}$,
assuming $\delta$ can be determined to good accuracy and $\beta$
is known or assumed. In that case, one finds
\begin{equation}
  \left\langle \delta_{\gamma'}^2\right\rangle = \alpha_g^{-2}
  \left\{\left\langle \delta_{M_{\rm E}}^2\right\rangle + 4
  \left\langle \delta_{\sigma_{||}}^2\right\rangle \right\},
\end{equation}
with 
$$\alpha_g \equiv \frac{1}{2}\left(\frac{\partial \log
  f}{\partial \log \gamma'} - \gamma' \, \log\left[\frac{R_{\rm
  A}}{R_{\rm E}}\right]\right),
$$ where $\delta_{\dots}$ indicate the fractional errors on the
respective quantities. Because in general $\delta_{M_{\rm E}}<
\delta_{\sigma_{||}}$, one finds the simple rule of thumb that the
error $\delta_{\gamma'} \sim \delta_{\sigma_{||}}$ for
close-to-isothermal mass models (typically one finds $\alpha_g
\sim 2$). We thus conclude that the fractional error on the determine
density slope $\gamma'$ is of the same order as the measurement error
on $\sigma_{||}$. This simple rule-of-thumb is confirmed for more
complex models as well (e.g.\ Koopmans et al.\ 2005) and typically
$\delta_{\gamma'} < 0.05$ for reasonable (spectroscopy) integration times
of $\sim$4\,hrs on 8--10\,m class telescopes for $L_*$ lens galaxies out to
$z\approx 1$.

\section{Observational Programs}

Two observational programs to combine gravitational lensing and
stellar kinematic constraints are described, that are currently
underway: (i) The Lenses Structure and Dynamics (LSD) Survey and (ii)
the Sloan Lens ACS (SLACS) Survey. I will mainly focus on their
results regarding the determination of galaxy density profiles and
their mass structure.

\subsection{The Lenses Structure \& Dynamics Survey -- LSD}

The Lenses Structure and Dynamics (LSD) Survey was started to
combine stellar-kinematic constraints with those from gravitational
lensing (e.g.\ Koopmans \& Treu 2003; Treu \& Koopmans 2004), to
extend E/S0 studies {\sl beyond} the local Universe.

Analysis of the five pressure supported E/S0 galaxies in our sample at
$z \approx 0.5-1.0$ was presented in Treu \& Koopmans (2004), with the
following results: (1) Extended dark-matter halos is detected in all
five E/S0 galaxies at $>99\%$~CL and dark matter contributes 40--70\%
(15--65\%) of the total mass within the Einstein (effective) radius.
(2) The inner power-law slope of the {\sl total} mass distribution is
$\langle \gamma'\rangle =1.75\pm0.10$ (with $\rho\propto
r^{-\gamma'}$), close to isothermal.  Including two more similarly
analysed systems (PG1115+080 and B1608+656) gives $\langle
\gamma'\rangle =1.9\pm0.1$ with 0.3 rms scatter, indicating that dark
and luminous mass already `conspire' at $z\approx 1$ to produce a
close-to-flat rotation curve, while preserving their spatial
segregation, similar to what is found in disk-dominated galaxies.  (3)
The inner power-law slope of the {\sl dark-matter} mass distribution
is $\langle \gamma\rangle =1.3^{+0.2}_{-0.4}$ (68\% CL), consistent
with numerical simulations for isotropic models with little adiabatic
contraction at low redshifts.
Although these results show that good-quality stellar kinematics of
E/S0 lens galaxies provide powerful constraints on the inner structure
of individual E/S0 galaxies out to $z\approx 1$, the study of trends
or evolution is clearly limited by the small sample size.

\begin{figure}[t]
   \centering
   \includegraphics[width=0.8\hsize]{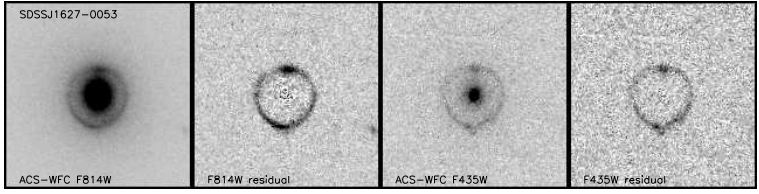}
      \caption{One of the newly discovered SLACS lens systems (Bolton
et al.\ 2005b)}
       \label{figure_lens}
   \end{figure}

\subsection{The Sloan Lens ACS Survey -- SLACS}

To increase the number of E/S0 galaxies suitable for lensing and
stellar dynamical studies, the Sloan Lens ACS (SLACS) Survey was
initiated (e.g.\ Bolton et al.\ 2005a\&b; Treu et al.\ 2005; Koopmans
et al.\ 2005). E/S0 lens candidates were initially selected from the
SDSS Luminous Red Galaxy (LRG) sample, based on the presence of
multiple emission lines at a redshift larger than that of the E/S0
galaxy (Bolton et al.\ 2004). The 49 candidates were ranked based on
their lensing probability and 39/49 were successfully targeted in a
cycle--13 HST snapshot survey through filters F435W and F814W. The
survey is continuing in cycle-14. Thus far a total of $>$25 new lens
systems have been discovered (e.g. Fig.1), making it the largest
sample of lens systems from a single well-defined survey.

The main results, thus far, presented in Bolton et al.\ (2005a\&b),
Treu et al.\ (2005) and Koopmans et al.\ (2005) are: (1) The SLACS
E/S0 lens galaxy sample appears in all respects (e.g.\ colours, FP) to
be a random sub-sample of the LRG sample, only skewed toward the brighter
(i.e.\ more massive) and higher surface brightness systems, because of
the lensing and lensing-cross-section selection effects. The SLACS
sample should therefore represents the general population of massive E/S0
galaxies. (2) The ratio between the central stellar velocity
dispersion ($\sigma$) and the velocity dispersion that best fits the
lensing models ($\sigma_{\rm SIE}$) is $\langle f_{\rm SIE}\rangle
\equiv \langle \sigma/\sigma_{\rm SIE}\rangle=1.01$ with 0.065 rms
scatter. (3) The average logarithmic density slope for the {\sl total}
mass density of $\langle \gamma' \rangle = 2.01^{+0.02}_{-0.03}$ (68\%
C.L.) inside $\langle {\rm R}_{\rm Einst} \rangle = 4.2 \pm
0.4$\,kpc. The inferred {\it intrinsic} rms spread in logarithmic
density slopes is $\sigma_{\gamma'}=0.12$. (4) The average
position-angle difference between the light distribution and the total
mass distribution is found to be $\langle \Delta \theta\rangle = 0 \pm
3$~degrees (rms of 10 degrees), setting an upper limit of $\langle
\gamma_{\rm ext}\rangle < 0.035$ on the average external shear.  The
total mass has an average ellipticity $\langle q_{\rm
SIE}\rangle$=0.78$\pm$0.03 (rms of 0.12), which correlates extremely
well with the stellar ellipticity, $q_*$, resulting in $\langle q_{\rm
SIE}/q_* \rangle = 0.99 \pm 0.03$ (rms of 0.11) for $\sigma >
225$~km\,s$^{-1}$. Assuming an oblate mass distribution and random
orientations, the distribution of ellipticities implies $\langle q_3
\rangle\equiv \langle(c/a)_\rho\rangle =0.66$ with an error of
$\sim$0.2.  (5) The average projected dark-matter mass fraction is
$\langle f_{\rm DM} \rangle =0.25 \pm 0.06$ (rms of 0.22) inside
$\langle {\rm R}_{\rm E}\rangle$, using the stellar mass-to-light
ratios derived from the Fundamental Plane as priors.  (6) Combined
with results from the {\sl Lenses Structure \& Dynamics} (LSD) Survey
at $z > 0.3$, we find no significant evolution of the total density
slope inside one effective radius for galaxies with $\sigma_{\rm
ap}\ge 200$~km\,s$^{-1}$ (Fig.2): a linear fit gives $\alpha_{\gamma'} \equiv
d\langle \gamma' \rangle/dz=0.23\pm0.16$ (1\,$\sigma$) for the range
$z$=0.08--1.01 .

\begin{figure}[t]
   \centering
   \includegraphics[width=0.8\hsize]{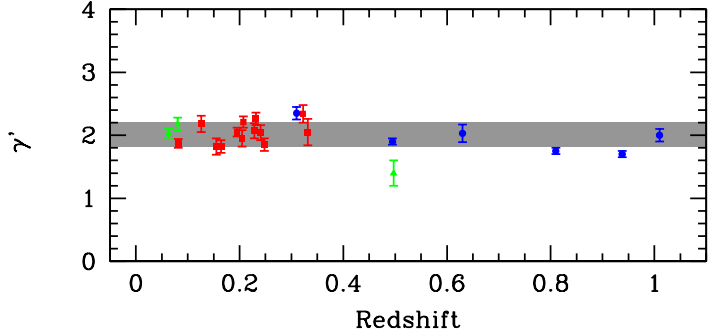}
      \caption{The density slope ($\gamma'$=$-d\log \rho/d\log r$) of
      SLACS E/S0 galaxies (red squares) and LSD E/S0 galaxies (blue
      circles, including PG1115+080 and CLASS B1608+656). The green
      triangles indicate all SLACS and LSD systems with
      $\sigma$$<$200\,km\,s$^{-1}$. The shaded region indicates
      the $68\%$ CL around the ensemble average slope $\langle \gamma'
      \rangle = 2.01$. Note the absence of significant evolution in
      the density slope.}
       \label{figure_lens}
   \end{figure}

\section{Conclusions}

The joint gravitational-lensing and stellar-kinematic analysis of E/S0
galaxies is a new, powerful, method that allows their density profiles
and mass structure (among other things) to be studied beyond the local
Universe. Two observational programs, the LSD and SLACS Surveys, have
shown very promising first results, quantifying the structure and
evolution of galaxies for the first time out to $z=1$. This ultimately
allows us to directly compare with and test the hierarchical
galaxy-formation scenario in the $\Lambda$CDM paradigm, based not on light 
but on mass.


\begin{thebibliography}{}

\bibitem[Bolton et al.(2004)]{2004AJ....127.1860B} Bolton, A.S., Burles, 
S., Schlegel, D.J., Eisenstein, D.J., Brinkmann, J.\ 2004, AJ, 127, 
1860 

\bibitem[Bolton et al.(2005)]{2005ApJ...624L..21B} Bolton, A.S., Burles, 
S., Koopmans, L.V.E., Treu, T., Moustakas, L.A.\ 2005a, ApJL, 624, 
L21 

\bibitem[]{} Bolton, A.S., Burles, S., Koopmans, L.V.E., Treu,
T., Moustakas L.A. 2005b, ApJ, accepted
 
\bibitem[]{}Kochanek, C.S., Schneider, P., Wambsganss, J., 2004, Part
2 of Gravitational Lensing: Strong, Weak \& Micro, Proceedings of the
33rd Saas-Fee Advanced Course, G. Meylan, P. Jetzer \& P. North,
eds. (Springer-Verlag: Berlin)

\bibitem[Koopmans \& Treu(2003)]{2003ApJ...583..606K} Koopmans, L.V.E., 
Treu, T.\ 2003, ApJ, 583, 606 

\bibitem[]{} Koopmans, L.V.E., Treu, T., Bolton, A.S., Burles,
S.C., Moustakas L.A. 2005, ApJ, submitted
 
\bibitem[Treu \& Koopmans(2004)]{2004ApJ...611..739T} Treu, T., 
Koopmans, L.V.E.\ 2004, ApJ, 611, 739 

\bibitem[]{} Treu, T., Koopmans, L.V.E., Bolton, A.S., Burles,
S.C., Moustakas L.A. 2005, ApJ, submitted 


\end{thebibliography}
\end{document}